\documentclass[12pt]{article}
\usepackage{epsfig}

\date{}
\begin{document}
\begin{flushright}
{FZJ-IKP(TH)-2000-28}
\end{flushright}
\vspace{1.0cm}

\begin{center}

{\Large \bf Mapping the proton unintegrated gluon distribution \\ in dijets correlations \\
in real and virtual photoproduction at HERA.

\vspace{1.0cm}}

{\large
A. Szczurek$^{1}$ }\\

\vspace{0.2cm}
$^{1}${ \em Institute of Nuclear Physics\\ PL-31-342 Cracow, Poland\\}

\vspace{0.6cm}

{\large %\bf 
N.N. Nikolaev$^{2,3}$, W. Sch\"afer$^{2}$ and  J. Speth$^{2}$}\\

\vspace{0.2cm}
$^{2}${ \em Institut  f\"ur
Kernphysik (Theorie), Forschungszentrum J\"ulich,\\ D-52425 J\"ulich, Germany\\}

$^{3}${ \em L.D.Landau Institute for Theoretical Physics\\ Chernogolovka,
Moscow Region 142 432, Russia}\\

\vspace{0.5cm}

\end{center}

\begin{abstract}
We discuss how
%the predictions of a recent two-component color dipole model for
 the dijet azimuthal correlations in DIS and
real photoproduction at HERA probe the differential (unintegrated)
gluon distribution in the proton.
%In contrast to early discussions in the literature
We find a strong
dependence of the azimuthal correlation pattern on Bjorken-$x$,
photon virtuality and the cut on the jet transverse momenta.
A rise of the azimuthal decorrelations is observed with decreasing
Bjorken-x due to the interplay of perturbative and nonperturbative effects.
%A stronger azimuthal decorrelation is predicted
%in photoproduction than in DIS.
%Rapidity gap between the two-jets decreases with increasing Bjorken-x.
We predict a strong rise of the same-side jet rate with photon energy for real
photoproduction. 
%A dedicated analysis of the experimental data could verify the
%two component picture and allow
We discuss conditions for the correlation function to be dominated by
hard perturbative gluons and
ways of constraining the size of the nonperturbative soft component.
We make some predictions for the THERA energy range.
The analysis of the energy dependence of the isolated jet and
two-jet cross sections in photoproduction 
would be a new way to study the not yet well constrained unintegrated
gluon distributions and to explore the onset of the pQCD regime. 
\end{abstract}

\newpage 

%---------------------
\section{Introduction}
%---------------------

Most of the recent studies of low-$x$ dynamics relevant for
HERA concentrated
on the analysis of the inclusive structure function, i.e. on the total
cross section for $\gamma^* p$ scattering \cite{total} and on diffraction
\cite{diffraction}. Because at low $x$ photon-gluon fusion is the dominant
underlying mechanism, such studies open the possibility of better understanding
the so-called unintegrated gluon distributions, the quantity first introduced
in \cite{BFKL}.

The jet studies are known to be a good tool to test perturbative
QCD effects.
It was pointed out already some time ago that dijet production
in DIS could be a method to study the onset of BFKL dynamics both in photo-
\cite{FR94} and electroproduction \cite{AGKM94}.
Unfortunately in practice, due to unavoidable cuts on transverse momenta of jets,
one samples rather large values of the gluon longitudinal momentum fraction $x_g$,
where it is not completely clear what is the underlying dynamics and
in particular what unintegrated gluon distribution should be used.
Up to now mostly real photoproduction data were published by ZEUS and H1
collaborations \cite{photo_data}.
Recently the H1 collaboration at HERA made available some new results
for jet production \cite{H1_jets}. However, in all these analyses only rather
inclusive observables have been discussed.

In the present note we discuss the jet production beyond the familiar
collinear approximation and focus on how more exclusive and
more differential jet production observables probe the unintegrated
gluon distribution. Based on the unintegrated gluon distributions
found recently \cite{IN00} from the phenomenological analysis of
$\sigma_{\gamma^*p}^{tot}$ we explore dijet azimuthal decorrelations.
%In this work the unintegrated gluon distribution was divided
%into a perturbative (hard) and nonperturbative (soft) regions of transverse
%momentum.
%Here we explore consequences of such a division for dijet azimuthal correlations.
In Ref.\cite{KMS99} such decorrelation effects were discussed in a perturbative
BFKL approach, and as a consequence only decorrelation effects at some distance
from the back-to-back configuration could be analyzed. The extension to
the whole azimuthal phase space requires understanding the differential
gluon densities in non-perturbative soft region.
The modelling of the nonperturbative soft gluon exchange as done in
Ref.\cite{IN00} allows us to extend the region of the azimuthal angle
between jets closer to $\pi$ and to explore the onset of the hard regime.  

%----------------------
\section{The formalism}
%----------------------

Here we collect basic formulae used throughout the present note.

\begin{figure} [H]
\begin{center}
\epsfig{file=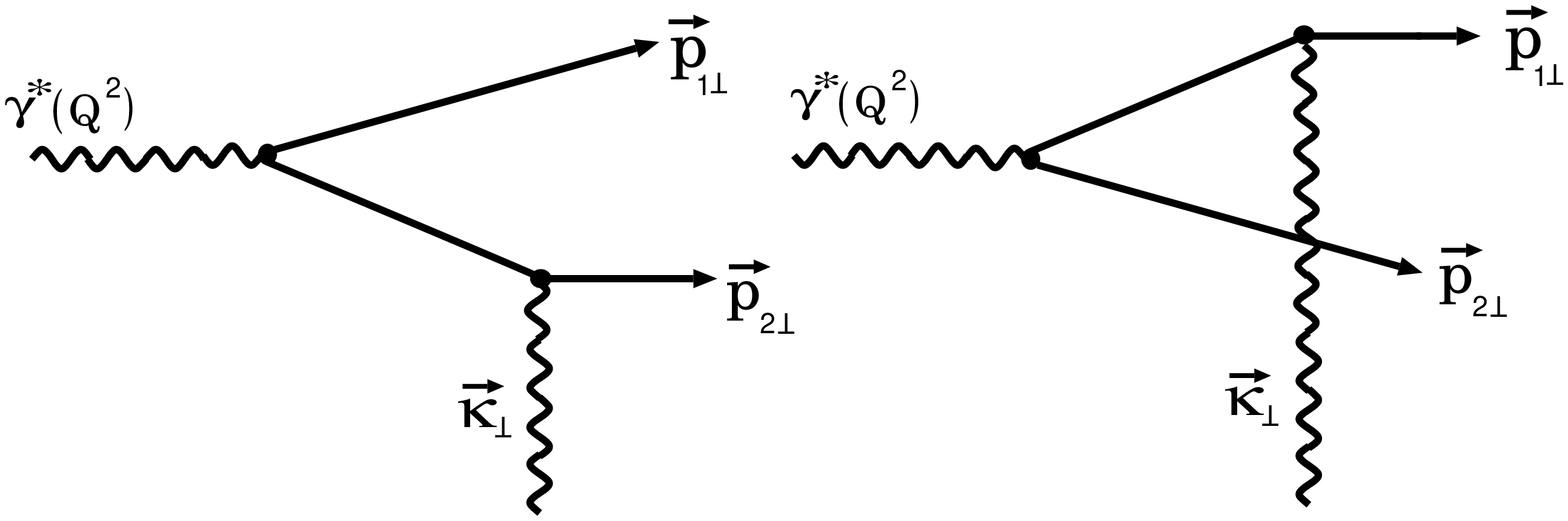, height = 16.0cm, width=7.0cm, angle=270}
\end {center}
\caption{Dijet production via photon-gluon fusion.
$\vec{\kappa_{\perp}}$ is the gluon transverse momentum.
}

\label{fig_diagram}
\end{figure}

At the parton level the total cross section for quark-antiquark
dijet production
$ \gamma^* + p \rightarrow j_1 + j_2 + X$ (see Fig.\ref{fig_diagram})
can be written in a compact way as:
\begin{equation}
\sigma_{T/L}^{\gamma^* p \rightarrow j_1 j_2}(x,Q^2) =
\int d\phi
 \int_{p_{1,\perp,min}^2} dp_{1,\perp}^2
 \int_{p_{2,\perp,min}^2} dp_{2,\perp}^2
\; \;
\frac{f_g(x_g,\kappa^2)}{\kappa^4} \cdot
 {\tilde \sigma}_{T/L}(x,Q^2,\vec{p}_{1,\perp},\vec{p}_{2,\perp}) \; ,
\label{total_dijets}
\end{equation}
where $x$ and $Q^2$ are standard kinematical variables.
In the formula above $f_g(x_g,\kappa^2)$ is the unintegrated gluon
distribution, which will be specified somewhat later, and
$\vec{\kappa}$ is the transverse momentum of the exchanged gluon.
It is related to the quark/antiquark jet transverse momenta
$\vec{p}_{1,\perp}$ and $\vec{p}_{2,\perp}$ as:
\begin{equation}
\vec{p}_{2,\perp}=\vec{\kappa} - \vec{p}_{1,\perp} \, , \, 
\kappa^2 = p_{1,\perp}^2 + p_{2,\perp}^2 + 2 p_{1,\perp} p_{2,\perp} cos \phi \; .
\end{equation}
We have written explicitly lower cuts on the transverse momenta
of jets in (\ref{total_dijets}).
The indices $T$ and $L$ refer to transverse and longitudinal photons,
respectively.
The auxilliary quantities introduced in (\ref{total_dijets})
\begin{eqnarray}
\tilde{\sigma}_T(x,Q^2,\vec{p}_{1,\perp},\vec{p}_{2,\perp})
 = \frac{\alpha_{em}}{2} \cdot \int dz \sum_f e_f^2 \alpha_s(l^2)
\nonumber \\
\left\{
[ z^2 + (1-z)^2 ]
 \left|
 \frac{\vec{p}_{1,\perp}}{p_{1,\perp}^2+\varepsilon_f^2} +
 \frac{\vec{p}_{2,\perp}}{p_{2,\perp}^2+\varepsilon_f^2} 
 \right|^2
 + m_f^2 \left(
 \frac{1}{p_{1,\perp}^2+\varepsilon_f^2} - 
 \frac{1}{p_{2,\perp}^2+\varepsilon_f^2}
  \right)^2
 \right\}
\label{aux_T}
\end{eqnarray}
for transverse photons and
\begin{equation}
\tilde{\sigma}_L(x,Q^2,\vec{p}_{1,\perp},\vec{p}_{2,\perp})
 = \frac{\alpha_{em}}{2} \cdot \int dz \sum_f e_f^2 \alpha_s(l^2)
\left\{
4 Q^2 z^2(1-z)^2
\left(
\frac{1}{p_{1,\perp}^2+\varepsilon_f^2} - \frac{1}{p_{2,\perp}^2+\varepsilon_f^2}
\right)^2
\right\}
\label{aux_L}
\end{equation}
for longitudinal photons.
Above we introduced
\begin{equation}
\varepsilon_f^2 = z(1-z)Q^2+m_f^2 \, .
\label{denominator}
\end{equation}
The unintegrated gluon distribution $f_g$ is evaluated at
\begin{equation}
x_g = \frac{M_t^2 + Q^2}{W^2 + Q^2} \;  ,
\label{x_g}
\end{equation}
where
\begin{equation}
M_t^2 = 
   \frac{p_{1,\perp}^2+m_f^2}{z}
 + \frac{p_{2,\perp}^2+m_f^2}{1-z}
\label{M_t2}
\end{equation}
is flavour dependent. It is obvious then that at large
transverse momenta of jets one samples larger values of $x_g$ than in
the case of total cross section.
The scale of the running coupling constant in (\ref{aux_T}) and (\ref{aux_L})
is taken to be $l^2 = \max(\kappa^2,\varepsilon_f^2+p_j^2)$, where
for small $l^2$ the coupling constant is 
frozen as in \cite{IN00}.

The gluon momentum $\kappa$ is responsible for the jets being not exactly
back-to-back in contrast to the conventional collinear approximation
to leading order.
In the following we limit ourselves to the region of $x_{\gamma} \sim$ 1,
where the jets are dominantly produced from the quark box on the very top
of the gluonic ladder.
%In all the examples discussed in the present note we have assumed that
%the jets are produced from the quark box on the very top of the gluonic
%ladder.
In doing so we restrict ourselves to leading order parton calculation and
omit jets from the ladder.
% They are interesting by themselves but are too complicated to be reliably
% calculated.
% However, by putting cuts on rapidities, limiting to the
% photon hemisphere, and/or on $x_{\gamma}$ \cite{BZ00}, limiting to direct jet
% production only $x_{\gamma}\sim$ 1, one could 
% study experimentally the effects discussed
% in the present paper.

The principal issue is how the isolated single jet and dijet production samples 
the unintegrated gluon distribution and when the azimuthal correlation function
will be dominated by hard perturbative gluons.
%Let us summarize now briefly the unintegrated gluon distribution used
%in the present analysis.
Although in the present note we address this question based on
the unintegrated gluon distribution
$f_g(x_g,\kappa^2)$ from a recent analysis in \cite{IN00} where it was
modelled phenomenologically to describe structure function data at
low Bjorken-$x$ and the total cross section for real $\gamma + p$ scattering,
we believe that our principal conclusions are to a great extent
model-independent.
The following simple two-component Ansatz was adopted in \cite{IN00}
\begin{equation}
f_g(x_g,\kappa^2) =
  {\cal F}_{soft}(\kappa^2) \frac{\kappa_s^2}{\kappa^2+\kappa_s^2}
+ {\cal F}_{hard}(x_g,\kappa^2) \frac{\kappa^2}{\kappa^2+\kappa_h^2}
\; .
\label{decomposition}
\end{equation}
The parameters $\kappa_s$ and $\kappa_h$ determine the scale of
the transition from the hard to soft gluon region \cite{IN00}.

The soft nonperturbative component was chosen in the Born form
\begin{equation}
{\cal F}_{soft}(\kappa^2) = a_{soft} C_F N_c \frac{\alpha_s}{\pi} V(\kappa^2)
\frac{\kappa^4}{(\kappa^2+\mu_{soft}^2)^2} \; .
\label{glue_soft}
\end{equation}
The vertex function $V(\kappa^2)$ is expressed by
the single-body isoscalar nucleon form factor
\begin{equation}
V(\kappa^2) \approx 1 - F(3\kappa^2) \; .
\label{vertex_function}
\end{equation}
The standard dipole parametrization for $F$ was used.

\begin{figure} [H]
\begin{center}
\epsfig{file=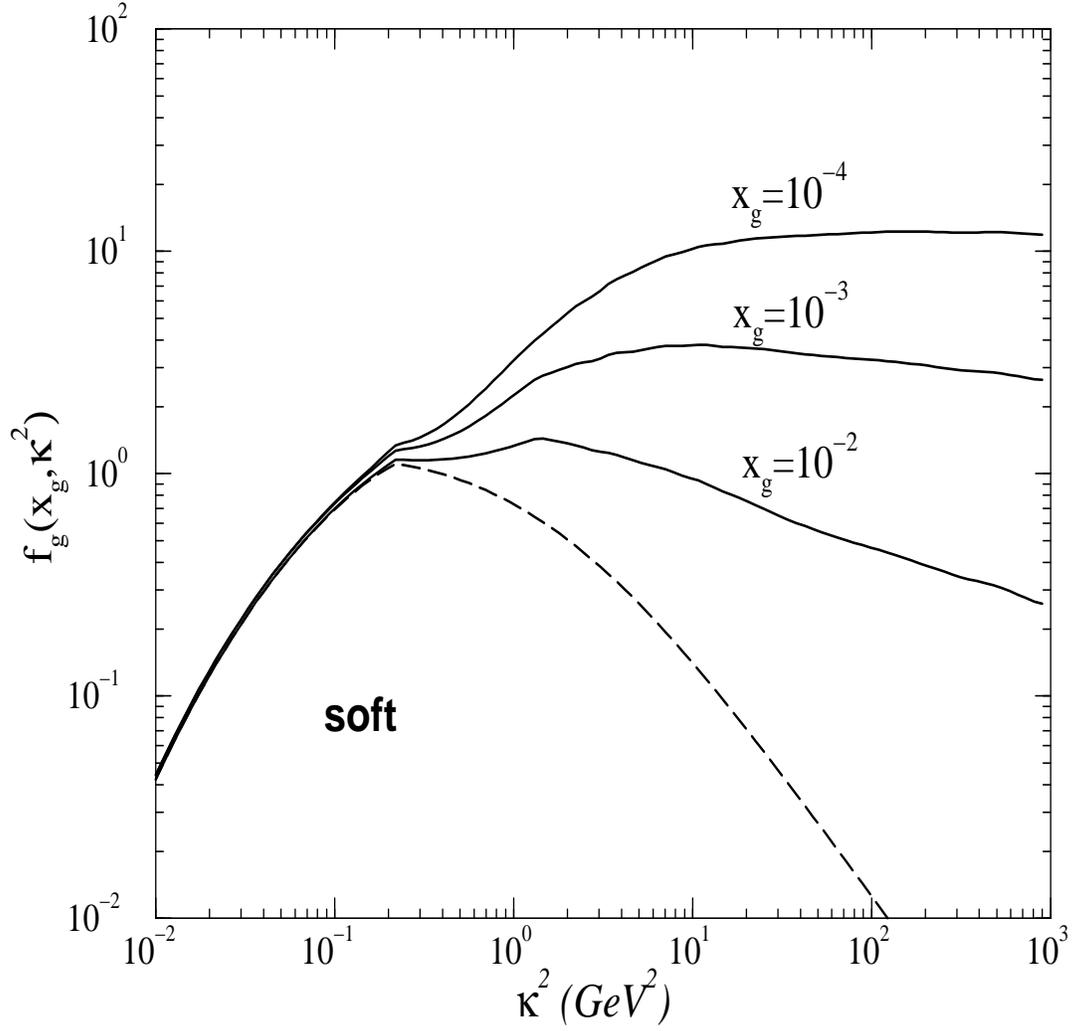, height = 14.0cm, width=14.0cm, angle=270}
\end {center}
\caption{The unintegrated gluon distribution as a function
$\kappa^{2}$, for different values of $x_g$.
}

\label{fig_glue}
\end{figure}

The hard component was taken in the form
\begin{equation}
{\cal F}_{hard}(x_g,\kappa^2) =
{\cal F}^{(B)}(\kappa^2) \frac{ {\cal F}_{pt}(x_g,Q_c^2)}{{\cal F}^{(B)}(Q_c^2)}
\theta(Q_c^2 - \kappa^2) + 
{\cal F}_{pt}(x_g,\kappa^2) \theta(\kappa^2 - Q_c^2) \; , 
\label{glue_hard)}
\end{equation}
where ${\cal F}^{(B)}$ is of the Born form (see \cite{IN00}).
The not yet specified ${\cal F}_{pt}$ is calculated from known conventional
DGLAP parametrizations as
\begin{equation}
{\cal F}_{pt}(x_g,\kappa^2) = \frac{\partial G_{DGLAP}(x_g,\kappa^2)}
{\partial \log \kappa^2 } \; .
\end{equation}
The results presented in this note were obtained based on a recent
MRST98 LO parametrization \cite{MRST98}.
As an example in Fig.\ref{fig_glue} we show the resulting unintegrated gluon
distribution as a function of $\kappa^2$ for a few different values of $x_g$.
The $x_g$-independent soft component is shown separately by the dashed line.
The model in \cite{IN00} is limited to small $x_g \ll$ 0.1 only, and should not
be applied for $x_g >$ 0.03.
%At $x_g >$ 0.1 the derivative becames negative which clearly demonstrates
%the limited range of applicability of the form given by Eq.(\ref{decomposition}). 
For other technical details we refer the reader to \cite{IN00}.

The two-component structure (\ref{decomposition}) of the unintegrated
gluon distribution as displayed in Fig.\ref{fig_glue}
leads to interesting consequences for the dijet azimuthal
correlations which will be studied in the following section.
The hard/soft decomposition of the gluon distribution found in \cite{IN00}
is sufficiently generic to believe that our major conclusions
on the onset of the hard regime are not strongly model dependent and are relevant
also to BFKL driven models.

%----------------
\section{Results}
%----------------

%------------------
\subsection{Dijets}
%------------------

In the present note we shall discuss mainly the effect of azimuthal jet-jet
correlations and leave the effect of other correlations for a separate analysis.
We shall limit ourselves to study the region of small Bjorken-$x$ only.
The cross section for the dijet production strongly depends on cuts
imposed on kinematical variables. In order to better demonstrate the effect of
coexistence of perturbative and nonperturbative effects
in the following analysis we shall restrict ourselves to cuts on kinematical variables
in the so-called hadronic center of mass (HCM) sytem ($\gamma^*$-proton center
of mass).
In the present purposefully simplified analysis we impose the cuts on
the parton level and avoid extra cuts in the laboratory frame. 

\begin{figure} [H]
\begin{center}
\epsfig{file=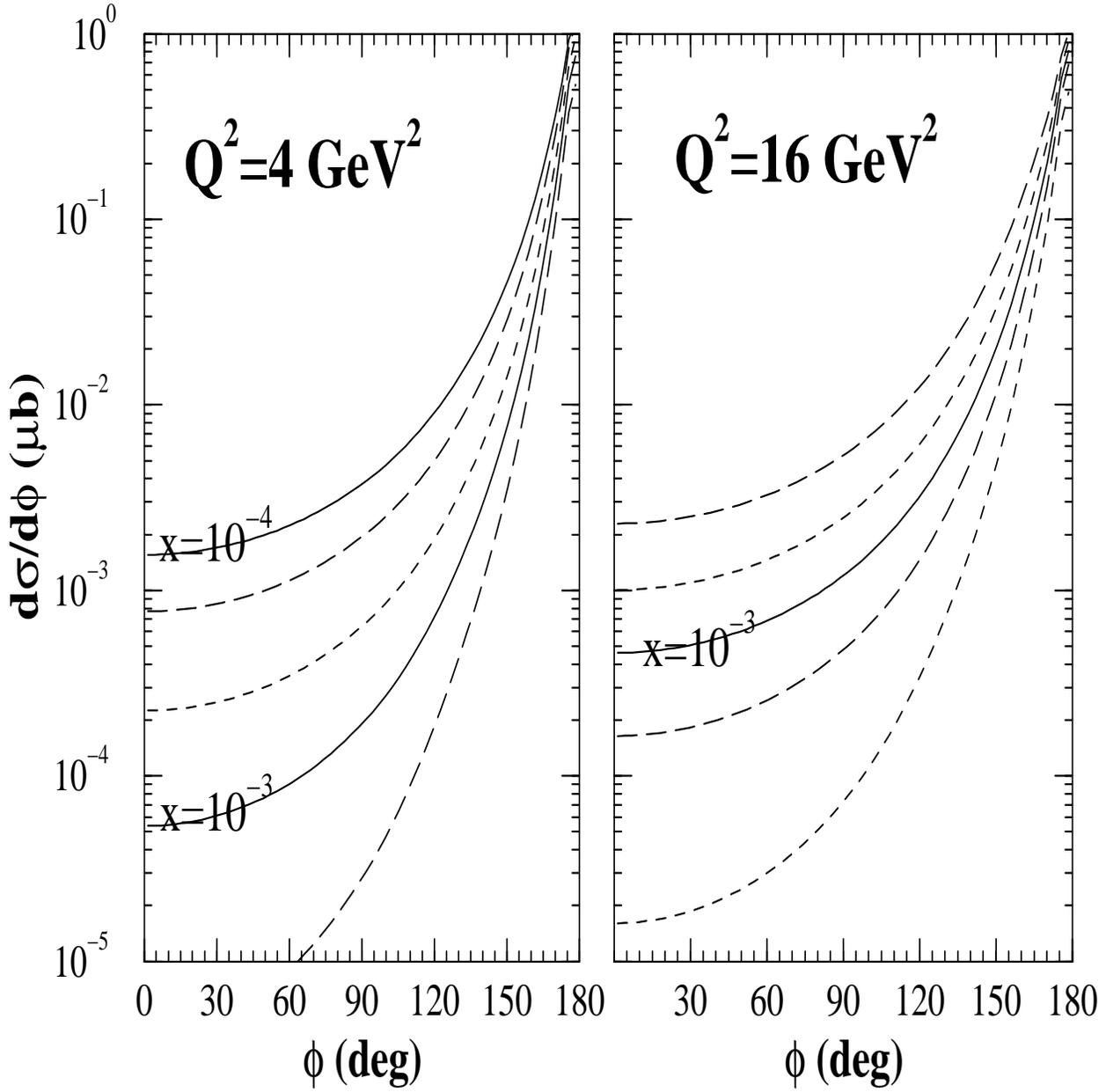, height = 14.0cm, width=14.0cm}
\end {center}
\vspace{2cm}
\caption{The cross section for $ \gamma^* p \rightarrow j_1 j_2 $ 
as a function of HCM azimuthal angle between jets for $Q^2$ = 4 GeV$^2$
(left panel) and $Q^2$ = 16 GeV$^2$ (right panel) for several values
of Bjorken-$x$ = 10$^{-4}$ (solid), 2 $\cdot$ 10$^{-4}$ (long-dashed),
5 $\cdot$ 10$^{-4}$ (short-dashed), 10$^{-3}$ (solid),
2 $\cdot$ 10$^{-3}$ (long-dashed),
5 $\cdot$ 10$^{-3}$ (short-dashed).
%, 10$^{-2}$ (solid).
}

\label{fig_x_dep}
\end{figure}

\begin{figure} [H]
\begin{center}
\epsfig{file=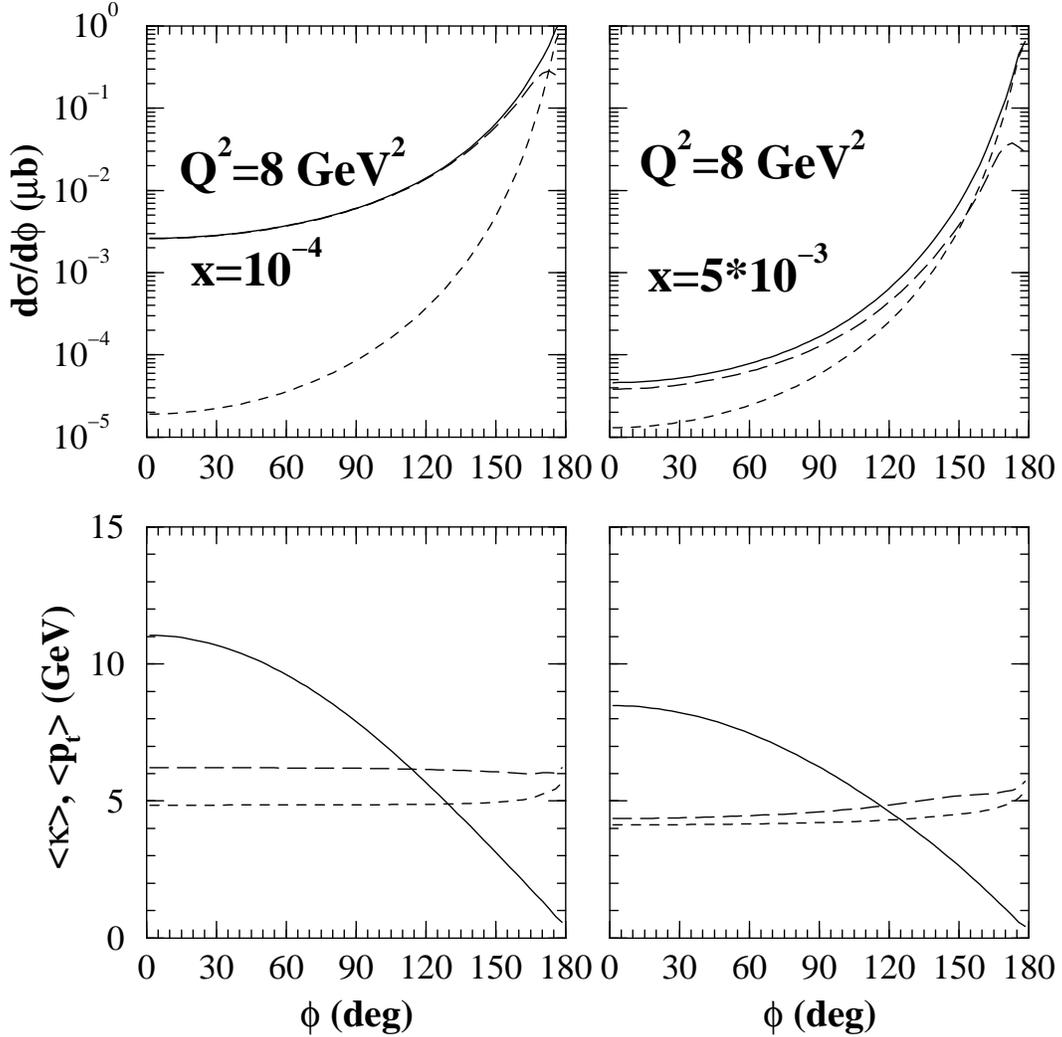, height = 14.0cm, width=14.0cm}
\end {center}
\vspace{2cm}
\caption{The decomposition of the HCM azimuthal correlation function $w(\phi)$
into hard (long-dashed) and soft (short-dashed) components for $x$ = 10$^{-4}$
(upper left panel) and $x$ = 5$\cdot$10$^{-3}$ (upper right panel). In this calculation
$Q^2$ = 8 GeV$^2$ and $p_{t,cut}$ = 4 GeV. In the lower panels, the
average gluon transverse momentum $<\kappa>$ (solid line) 
and $p_{\perp,soft}$ (short-dashed line) and $p_{\perp,hard}$ 
(long-dashed line) are shown correspondingly.} 

\label{fig_x_deco}
\end{figure}

In Fig.\ref{fig_x_dep}, we present $d \sigma(\gamma^*p \rightarrow j_1 j_2)/d \phi$
as a function of HCM azimuthal angle between jets for two different values
of photon virtuality $Q^2$ = 4 GeV$^2$ (left panel) and
$Q^2$ = 16 GeV$^2$ (right panel) for a series of Bjorken $x$.
In this calculation, we have restricted the transverse
momenta of jets to $p_{1,\perp}^{HCM},p_{2,\perp}^{HCM} > p_{t,cut} =$ 4 GeV
and summed over light flavours $u$, $d$ and $s$. While in the case of
the total cross section the effective mass $m_f$ of the quark is responsible
for confinement effects, in the dijet production the cross section
is in practice independent of the explicit value of $m_f$.
One can observe a strong dependence of the azimuthal angle decorrelation
pattern on Bjorken $x$.
A closer inspection of both panels simultaneously leads to the conclusion
that averaging over a broad range of $Q^2$ would to a large extent destroy the
effect as it involves automatically averaging over a certain range of $x_g$,
the most crucial variable for the effect to be observed.
 A significant part of the effect is due to the
interplay of the hard(perturbative) and soft(nonperturbative) components.
This is explained better in Fig.\ref{fig_x_deco} where
the correlation function
% defined as:
%
%\begin{equation}
%w(\phi) = \frac{d \sigma(\gamma^* + p \rightarrow j_1 + j_2) / d \phi} 
%{ \sigma(\gamma^* + p \rightarrow j_1 + j_2) } 
%\end{equation}     
%
is decomposed into hard and soft components
for two rather different values of Bjorken-$x$: $x$ = 10$^{-4}$ (left panel)
and $x$ = 5 $\cdot$ 10$^{-3}$ (right panel).
In this calculation the virtuality of the photon was fixed to $Q^2$ = 8 GeV$^2$. 
While at $x$ = 10$^{-4}$ the hard component dominates,
at larger $x$ the soft component becomes equally important.
%The effect of the interplay between soft and hard effects on the azimuthal
%correlation may be misinterpreted as a genuine BFKL effect.
%In a perturbative BFKL approach the weakening of the azimuthal back-to-back
%correlation is purely due to diffusion in $\kappa$ \cite{AGKM94}.
In the region of small $\phi$ and small $x$ our results are similar to
those from perturbative BFKL dynamics \cite{AGKM94}, which is rather a general
feature of the $k_{\perp}$-factorization approach.

The onset of the hard regime can be best seen in the lower part of
Fig.\ref{fig_x_deco} where we present averaged
value of the gluon transverse momentum $\kappa$ (solid line) as sampled
at different azimuthal angle $\phi$ between jets. For comparison we
also show average transverse momentum
of the harder ($p_{\perp,hard} = \max(p_{1,\perp},p_{2,\perp})$, long-dashed line)
   and softer ($p_{\perp,soft} = \min(p_{1,\perp},p_{2,\perp})$, short-dashed line)
 quark(antiquark) jet.
The region of small azimuthal angle $\phi$
(same-side jets) samples on average rather large values of gluon transverse
momentum $\kappa$ where the perturbative component dominates.
Here on average gluon transverse momentum is larger than the transverse
momentum of each of the two jets.
At $\phi$ = 0, $<\kappa> = <p_{\perp,hard}> + <p_{\perp,soft}>$.
In our case ($x_{\gamma} \sim$ 1) the momentum of the same-side jets $\vec{\kappa}$
is compensated by the transverse momentum of hadrons (minijets) which carry
a very small fraction of photon's momentum.
Close to $\phi = \pi$ the average $\kappa$ is rather small, smaller
than the transverse momentum of each of the jets. In this region, models for
the soft component can be tested.

In the case of symmetric cut in a sharp (in $\kappa^2$) transition between
soft and hard region, to a crude approximation, the hard region can
be estimated as $\cos\phi > -1 + \kappa_0^2/2p_{t,cut}^2$, where $\kappa_0$ is
the transition value in the gluon transverse momentum.
In the model in \cite{IN00}, the border is not as well defined and
the transition region depends in addition on $x_g$.  
We wish to mention that in general the region of larger $\phi$
($|\phi - \pi| \sim$ 0) is model dependent.
In our model, the cross section depends on the treatment of
the infra-red region i.e. the formula used for $\alpha_s$ and the choice of
its argument.
We hope that studying this region experimentally could provide new information
about genuine nonperturbative effects which are rather poorly known up to now.

\begin{figure} [H]
\begin{center}
\epsfig{file=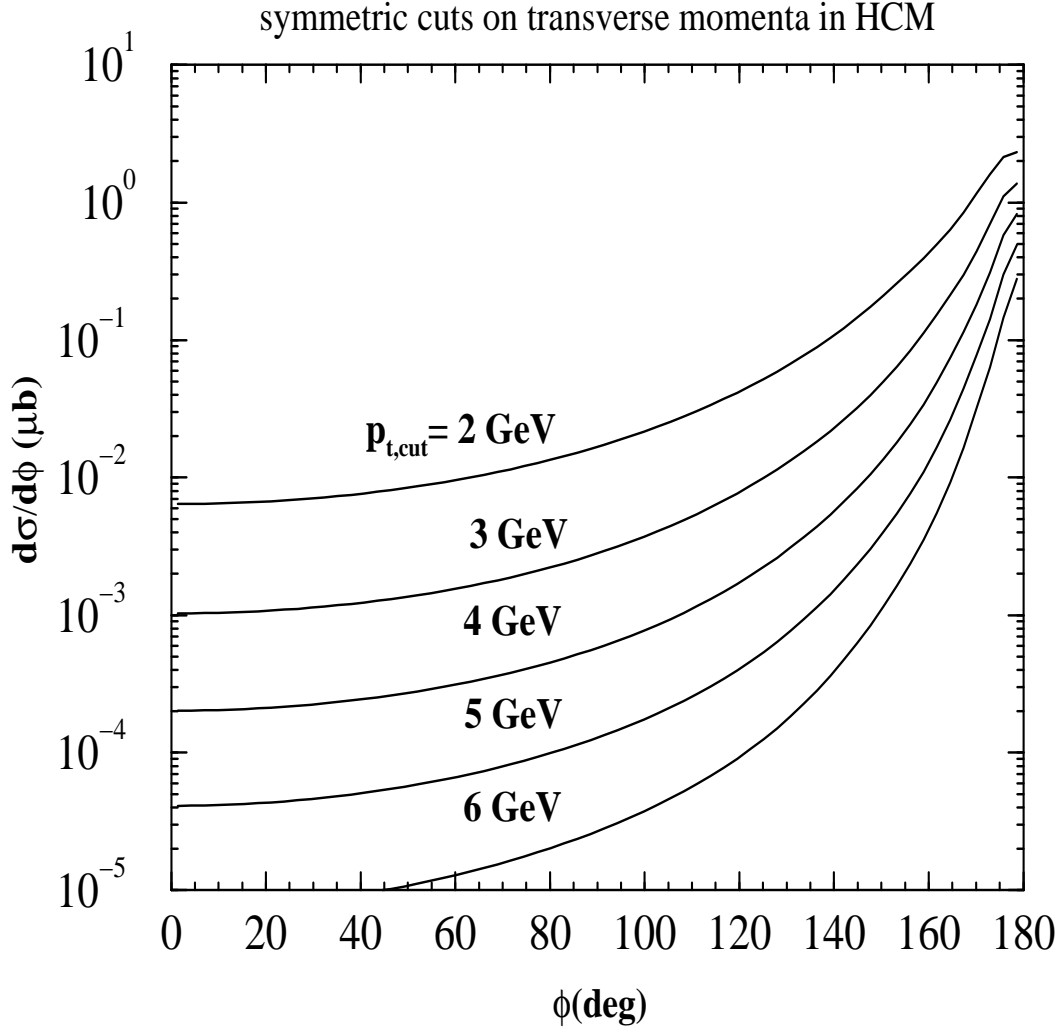, height = 14.0cm, width=14.0cm, angle=270}
\end {center}
\vspace{2cm}
\caption{
The cross section for $\gamma^* p \rightarrow q \bar q$ as a function
of HCM azimuthal angle between jets for $Q^2$ = 8 GeV$^2$, $x$ = 10$^{-3}$
for several values of the HCM symmetric cut on the jet transverse momenta:
$p_{t,cut}$ = 2,3,4,5,6 GeV.
}

\label{fig_cut_dep}
\end{figure}

\begin{figure} [H]
\begin{center}
\epsfig{file=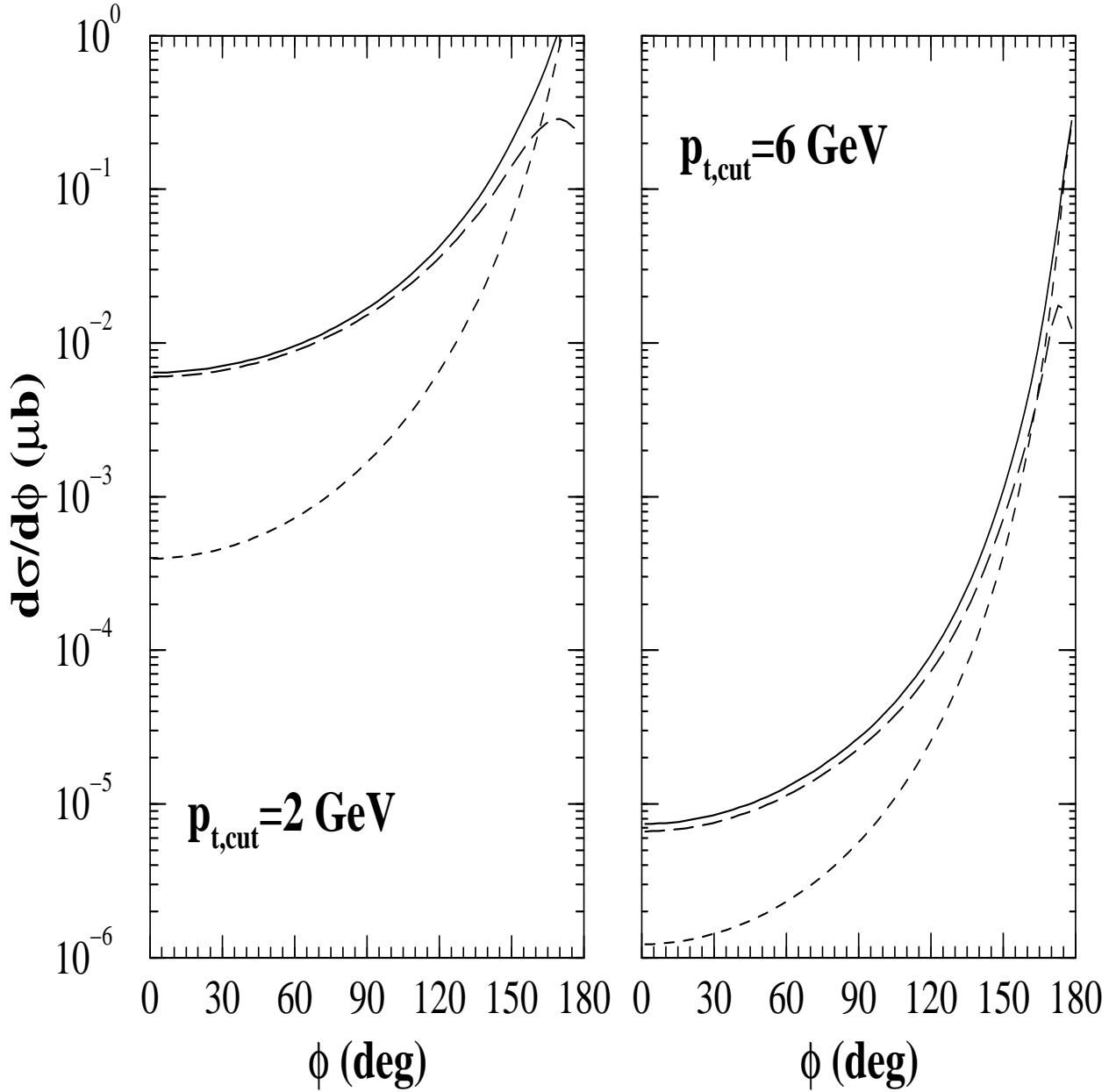, height = 14.0cm, width=14.0cm}
\end {center}
\vspace{2cm}
\caption{
The decomposition of the HCM azimuthal correlation function $w(\phi)$
into hard (long-dashed) and soft (short-dashed) components for
$p_{t,cut}$ = 2 GeV (left panel) and $p_{t,cut}$ = 6 GeV (right panel).
In this calculation $Q^2$ = 8 GeV$^2$ and $x$ = 10$^{-3}$.
}

\label{fig_cut_deco}
\end{figure}

The cut on the transverse momentum is usually a tool to define jets.
Studying the dependence of the result on the cut may however provide
some new information.
As an example in Fig.\ref{fig_cut_dep} we show the dependence
of $d \sigma /d \phi$
on the cut on the HCM transverse momenta of jets for fixed value of
$x$ = 10$^{-3}$ and $Q^2$ = 8 GeV$^2$ and in Fig.\ref{fig_cut_deco}
the corresponding decomposition 
%of the correlation function 
into soft and hard components
for $p_{t,cut}$ = 2 GeV (left panel) and $p_{t,cut}$ = 6 GeV (right panel). 
The small $p_{t,cut}$ = 2 GeV may be slightly academic but is chosen here
to better emphasize the effect.
In both cases we observe the dominance of the hard component at small $\phi$
and the soft component near to the back-to-back configuration.
In the traditional collinear approximation such a separation of soft and hard
component is not possible as they are both lumped together into the integrated
gluon density. 
In the collinear approximation in first order in $\alpha_s$, jets
are produced back-to-back and only higher order
corrections lead to an azimuthal decorrelation.
By comparing the two panels in Fig.4 one can observe
stronger back-to-back correlations for larger $p_{t,cut}$.
This is due to the fact that the larger $p_t$ region is unavoidably related
to larger $x_g$, that is to the region of $f_g$ in which the soft component is
more prominent.

The experimental identification of the effects discussed here requires 
good statistics in the data sample.
Up to now, in practice \cite{Maciej}, one averages rather over broader range of
Bjorken $x$, photon virtuality and jet transverse momenta.
Most of the effects discussed in the present note are then washed out and
the information about the small-$x$ dynamics is to a large extent lost.
It seems rather difficult at present to study the whole correlation function
$w(\phi)$ for fixed values of $x$ and $Q^2$.
We suggest that at present, instead of analyzing $w(\phi)$ itself,
one could study the ratio:
\begin{equation}
S(x,Q^2) \equiv 
 \frac{\displaystyle\int_{0}^{\pi/2} w(\phi;x,Q^2,p_{t,cut}) \; d\phi }
      {\displaystyle\int_{0}^{\pi}   w(\phi;x,Q^2,p_{t,cut}) \; d\phi }   \; ,
\label{same_side_ratio}
\end{equation}
i.e. the percentage of the same-side dijets of all dijets, as a function
of Bjorken-$x$, photon virtuality $Q^2$ and/or the cuts on transverse momenta
of jets $p_{t,cut}$.
In Table 1, we show as an example the predictions of the two-component
model \cite{IN00} for $S$ as a function of Bjorken-$x$ and $p_{t,cut}$ = 4 GeV.
Here the virtuality of the photon was fixed at $Q^2$ = 8 GeV$^2$.
As seen from the table, the two-component model predicts a significant
dependence of the same-side jet fraction $S$ on Bjorken-$x$ and the value
of the lower cut-off on jet transverse momenta in HCM.
The larger Bjorken $x$, the lower $S$ and the smaller $p_{t,cut}$, the larger $S$.
In fact these two effects are at least partially correlated.
It should be noticed that the larger $p_{t,cut}$ means automatically the larger
$x_g$ (see Eq.(\ref{x_g})).
These strong effects predicted here should
be easy to observe experimentally. However, a more detailed comparison
with experimental data should include the separation of jets in
rapidity/azimuthal angle space \footnote{This is more important for the
same-side jets.} and/or hadronization effects.
%Thus we do not expect that the numbers from the table will be exactly
%reproduced in a future experiment. We do expect, however, that the tendency
%predicted here should be visible.
The higher-order QCD corrections are expected to
increase $S$, especially for larger values of $x$ and/or larger
cuts on transverse momenta, i.e. in the region where it is small.
%In addition, the exact strength of the soft component may be not well
%determined from the inclusive data.  

\begin{figure} [H]
\begin{center}
\epsfig{file=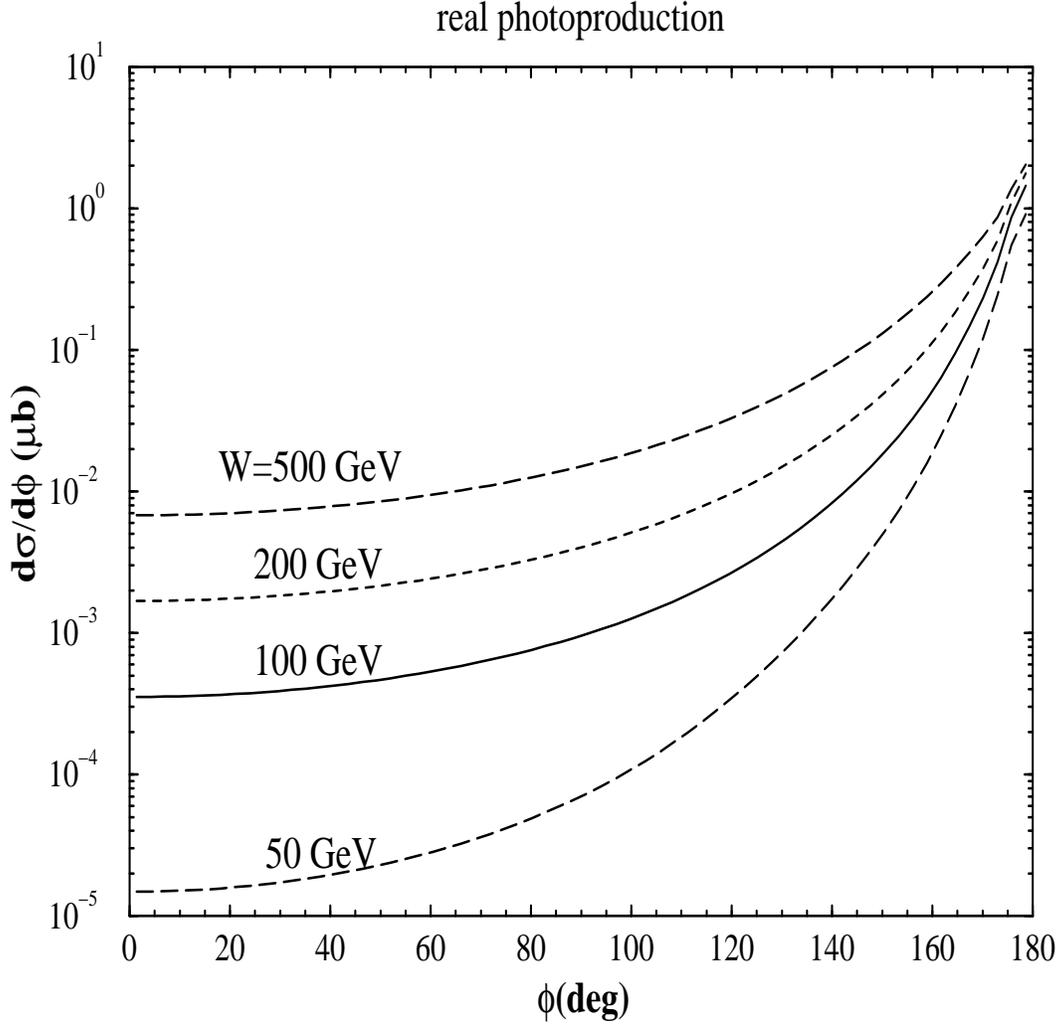, height = 14.0cm, width=14.0cm, angle=270}
\end {center}
\vspace{2cm}
\caption{
The correlation function of dijets in real photoproduction
for different $\gamma p$ center-of-mass energies
$W$ = 50, 100, 200, 500 GeV.
In this calculation a lower cut on both jets momenta $p_{t,cut} =$ 4 GeV
was imposed.
}

\label{fig_photo}
\end{figure}

In the model in Ref.\cite{IN00} the total (real) photoproduction cross section
at energies $W <$ 100 GeV is dominated by the soft component. Only at very high,
not yet available, energies the hard component would dominate. At "intermediate"
energy available at HERA, the two components coexist and their fraction is
a smooth function of initial $\gamma p$ energy.
In principle, the same stays true for the dijet production and
has interesting consequences for the jet azimuthal correlations.
In Fig.\ref{fig_photo}, we show the azimuthal correlation 
for a few different values of $\gamma p$ center of mass energy
$W$ = 50, 100, 200, 500 GeV. We observe strongly rising decorrelation
of jets when going from fixed target energies to energies relevant
at THERA \cite{THERA}.
Similarly as in the jet electroproduction, the key variable
here is the gluon momentum fraction, which is
$<x_g> \approx$ 5$\cdot$10$^{-2}$ at W = 50 GeV and
$<x_g> \approx$ 10$^{-3}$ at the THERA energy of 500 GeV.
We get $S \approx$ 0.5\% at W = 50 GeV.
At the THERA energy of 500 GeV we predict about 3.5\% of the same side jets.  
%In comparison to electroproduction case discussed above,
%in real photoproduction we observe significantly stronger back-to-back
%correlation of jets.
%A slowly increasing decorrelation of jets with increasing entrance energy can be seen
%from the figure. The significant difference between azimuthal correlation of jets
%in electro- and photoproduction is therefore another interesting prediction
%of the model in \cite{IN00} which could be studied experimentally.
The rise of the decorrelation with the photon energy in real photoproduction
%is another interesting prediction of the model in \cite{IN00} which
%could be studied experimentally in the future. It
 can be traced back to
larger values of $\kappa$ sampled by the same-side jets than by
the opposite-side jets. We note that the rise of the decorrelation
with increasing  energy obtained here based on the model from \cite{IN00},
is rather typical of $k_{\perp}$-factorization approach in general.
Therefore the experimental confirmation of this effect would be
a valuable test of this kind of approach. 

%--------------------------------------------------
\subsection{Isolated jets in the photon hemisphere}
%--------------------------------------------------

\begin{figure} [H]
\begin{center}
\epsfig{file=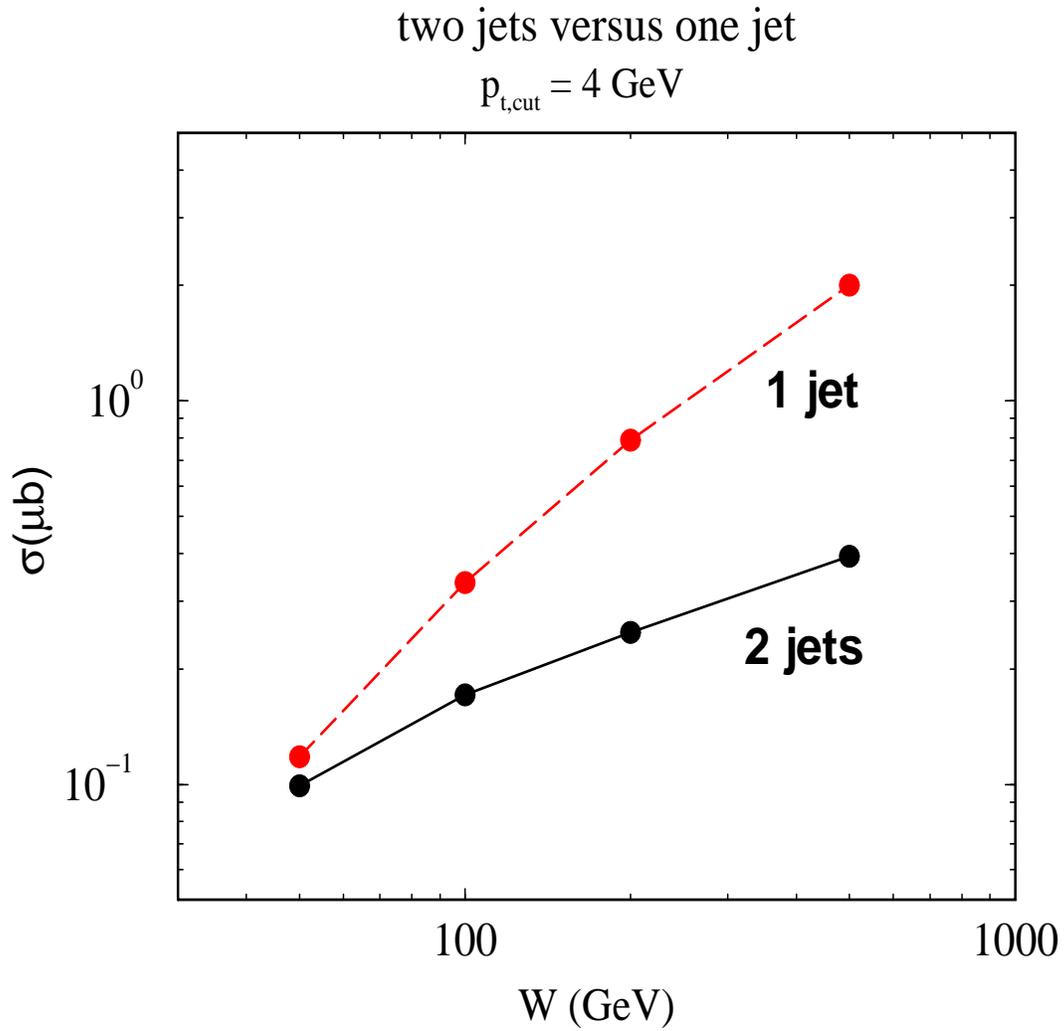, height = 14.0cm, width=14.0cm}
\end {center}

\vspace{2cm}
\caption{
The cross section for $\gamma p \rightarrow$ "two jet" (solid) and "one jet"
(dashed) cases as a function of $\gamma p$ CM energy. In this calculation
$p_{t,cut}$ = 4 GeV. 
}

\label{fig_1vs2}
\end{figure}

There is another interesting prediction of the two-component model
which we wish to discuss briefly now. Up to now we have discussed only cases
when \underline{both} jets have transverse momenta larger than a certain
cut value $p_{t,cut}$.
In principle there are also cases (events) with one hard ($p_{\perp} > p_{t,cut}$)
jet and one soft ($p_{\perp} < p_{t,cut}$) "jet".
In this single jet event $x_{\gamma} <$ 1, because the transverse momentum
of the single quark(antiquark) jet is compensated by a transverse momentum
of a much softer gluon.
\footnote{The dijets quark-gluon and antiquark-gluon can be separated out
by cuts on $x_{\gamma}$ and/or cuts on rapidities.} 
In the following we shall call such cases
"one jet" events for simplicity, as opposed to the previous case called here
"two jet" events for brevity. Let us compare the rate
of such "one jet" events to the previously discussed cases of two hard jets
in photoproduction.
As an example in Fig.\ref{fig_1vs2} we compare the cross section for the
two cases with the lower cut on transverse momentum $p_{t,cut}$ = 4 GeV.
Firstly, we observe that the cross section for both cases
are of similar order.
Furthermore we observe a significantly stronger rise of the cross section
for the "one jet" case than for the "two jet" case which may be
a bit surprising at least at first look.
This different energy dependence is related to
different $x_g$ and $\kappa$ sampled in both cases.
For example for W = 100 GeV and $p_{t,cut}$ = 4 GeV 
in the "one jet" case $<x_g> \approx$ 0.01 is substantially lower than
in the "two jet" case $<x_g> \approx$ 0.02. Both numbers are, however,
substantially larger than average $<x_g> \approx$ 0.005 sampled in the case
of total cross section. 
The effect of $\kappa$ is more complicated as averaged $\kappa$ strongly
depends on $\phi$ for the "two jet" case. The interplay of the two
effects ($x_g$ and $\kappa$) causes the unintegrated gluon distributions
to be sampled differently in the "one jet" and "two jet" cases. This,
at least potentially, allows the possibility of a further nontrivial test
of unintegrated gluon distributions. 
It would be valuable to compare the present predictions with
the predictions of standard (collinear) NLO approach
 (see for instance \cite{collinear}). 

%In all the examples discussed in the present note we have assumed that
%the jets are produced from the quark box on the very top of the gluonic
%ladder. This means we have omitted jets from the ladder.
%They are interesting by itself but are more complicated to be reliably
%calculated. However, by putting cuts on rapidities, limiting to the
%photon hemisphere, and/or on $x_{\gamma}$ \cite{BZ00}, limiting to direct jet
%production only, one could get rid of them and study the effects discussed
%in the present paper.

%--------------------
\section{Conclusions}
%--------------------

Based on the recent model determination \cite{IN00} of the unintegrated
gluon distribution in the proton we have explored the impact of the soft
gluon component and the onset of the perturbative regime on
%We have found that a recent model \cite{IN00} with a significant
%soft-gluon component in addition to the hard perturbative component
%in the unintegrated gluon distribution gives interesting predictions for
the dijet azimuthal correlations.
% In particular, this model division into hard
%and soft components leads to
We have predicted  a strong dependence of the azimuthal correlation pattern
on Bjorken $x$, photon virtuality and the cut on the jet transverse momenta.
The effects in the electroproduction could be verified now at HERA, provided
a careful differential ($x$ ,$Q^2$, transverse momentum cut) studies
of the dijets are made.

In real photoproduction we have predicted the rise of
the same-side jet rate with the photon energy. It would be important to
compare the results of the model discussed here with the result
of the standard collinear approach to understand the potential of such a dijet
study to shed more light on the low-$x$ dynamics which has been studied up to
now in rather inclusive processes. Finally, we have found that the study of
the energy dependence of the "one jet" (defined in the text) cross section
would be a new test of unintegrated gluon distributions.

\vskip 0.5cm

{\bf Acknowledgments}
One of us (A.S.) is indebted to Maciej Przybycie\'n for a discussion of
the ZEUS experimental results, Anna Sta\'sto for an exchange of information
about some details of their calculation \cite{KMS99} and Igor Ivanov for assistance
in using his routine for $f_g$ \cite{IN00}.
This work was partially supported by the German-Polish DLR exchange program,
grant number POL-028-98.

%-------------------------------------------------------------------------

%----------------------------------------------------------------------

\newpage

%{\bf FIGURE CAPTIONS}

%----------------------------------------------------------------------
\newpage

%------------------------------------------------------------------------

%-------------------------------------------------------------------------

%------------------------------------------------------------------------

%---------------------------------------------------------------------------

%---------------------------------------------------------------------------

%----------------------------------------------------------------------------

%----------------------------------------------------------------------------

%----------------------------------------------------------------------------

\begin{table}
\begin{tabular}{|l|c|c|c|}
\hline
 & \multicolumn{3}{c|}{ $p_{t,cut} (GeV)$ }                      \\
\hline 
Bjorken-x &   2   &   4   &   6   \\
\hline
$1 \cdot 10^{-4}$ & 6.41(-0) & 2.30(-0) & 9.35(-1)  \\
$2 \cdot 10^{-4}$ & 5.07(-0) & 1.58(-0) & 5.66(-1)  \\
$5 \cdot 10^{-4}$ & 3.42(-0) & 8.13(-1) & 2.23(-1)  \\
$1 \cdot 10^{-3}$ & 2.34(-0) & 4.00(-1) & 6.68(-2)  \\ 
$2 \cdot 10^{-3}$ & 1.46(-0) & 1.34(-1) & 2.54(-3)  \\
$5 \cdot 10^{-3}$ & 6.72(-1) & 5.83(-3) & --------  \\
$1 \cdot 10^{-2}$ & 3.15(-1) &  ------  & --------  \\
\hline
\end{tabular}  

\caption{The fraction of the HCM same-side jets $S(x,Q^2,p_{t,cut})$ in \% as
the function of Bjorken-$x$ and the symmetric cut on the jet
transverse momenta in GeV for $Q^2$ = 8 GeV$^2$.}

\end{table}

\end{document}